\documentclass[a4paper,11pt]{article}
\usepackage{cite}
\usepackage{microtype}
\usepackage{amsthm,amssymb,amsmath}
\usepackage{enumitem}
\usepackage[margin=30mm,includefoot,foot=12mm]{geometry}
\usepackage{hyperref}
\hypersetup{
    pdftitle={Distributed Colour Reduction Revisited},
    pdfauthor={Jukka Kohonen, Janne H. Korhonen, Christopher Purcell, Jukka Suomela, Przemyslaw Uznanski}
}

\newenvironment{mycover}
               {\list{}{\listparindent 0pt
                        \itemindent    \listparindent
                        \leftmargin    0cm
                        \rightmargin   0cm
                        \parsep        0pt}%
                \raggedright
                \item\relax}
               {\endlist}
               
\newcommand{\myemail}[1]{\,$\cdot$\, #1}
\newcommand{\myaff}[1]{\,$\cdot$\, #1\par\medskip}

\begin{document}

\begin{mycover}
{\huge\bfseries Distributed Colour Reduction Revisited \par}

\vspace{10mm}

\textbf{Jukka Kohonen}
\myemail{jukka.kohonen@cs.helsinki.fi}
\myaff{University of Helsinki}

\textbf{Janne H.\ Korhonen}
\myemail{janne.h.korhonen@aalto.fi}
\myaff{Aalto University}

\textbf{Christopher Purcell}
\myemail{christopher.purcell@aalto.fi}
\myaff{Aalto University}

\textbf{Jukka Suomela}
\myemail{jukka.suomela@aalto.fi}
\myaff{Aalto University}

\textbf{Przemys\l{}aw Uzna\'nski}
\myemail{przemyslaw.uznanski@inf.ethz.ch}
\myaff{ETH Z\"urich}
\end{mycover}

\vspace{-1mm}
\paragraph{Abstract.}
We give a new, simple distributed algorithm for graph colouring in paths and cycles. Our algorithm is fast and self-contained, it does not need any globally consistent orientation, and it reduces the number of colours from $10^{100}$ to $3$ in three iterations.

\vspace{6mm}

\newcommand{\cto}{\mathbin{\vartriangleright}}
\newcommand{\cF}{\mathcal{F}}
\newcommand{\cL}{\mathcal{L}}
\newcommand{\cX}{\mathcal{X}}
\newcommand{\cY}{\mathcal{Y}}
\newcommand{\cZ}{\mathcal{Z}}
\newcommand{\ccA}{\mathsf{A}}

\section{Introduction}

We present a very fast and simple distributed algorithm for colouring paths and cycles. Algorithms for colouring paths or cycles are key primitives that are used as subroutines in many other distributed and parallel algorithms, and they are also material that is typically covered in introductory courses on distributed algorithms \cite{barenboim13distributed,suomela12course,peleg00distributed,wattenhofer13lecture}. Yet the best currently known algorithms tend to be inefficient, restricted, or inelegant.

\subsection{Problem setting}

We will focus on \emph{iterative colour reduction algorithms} in paths and cycles: we have a path that is properly coloured with $k$ colours, and the algorithm will relabel the nodes so that the path is properly coloured with $f(k) < k$ colours. Naturally, we can then iterate the algorithm to find a colouring with $f(f(k))$ colours, $f(f(f(k)))$ colours, etc., until we reach a fixed point. We use the notion $k \cto f(k)$ for colour reduction from $k$ to $f(k)$.

We are interested in \emph{one-round algorithms} \cite{kuhn06complexity}, in which all nodes are re-coloured based on the current colours of their immediate neighbours only. Such an algorithm can be interpreted as a function
\[
    A\colon [k] \times [k] \times [k] \to [c]
\]
that maps three old colours to a new colour. Here $A(x,y,z)$ is the new colour of a node that was coloured $y$, and its two neighbours had colours $x$ and $z$. We want to keep the algorithm symmetric so that we can apply it even if we do not have a well-defined global orientation:
\begin{equation}
    A(x_1,x_2,x_3) = A(x_3,x_2,x_1) \text{ for all } x_1,x_2,x_3. \label{eq:sym}
\end{equation}
And naturally we require that the algorithm produces a proper colouring as output, assuming we have a proper colouring as input:
\begin{equation}
    A(x_1,x_2,x_3) \ne A(x_2,x_3,x_4) \text{ for all } x_1 \ne x_2 \ne x_3 \ne x_4. \label{eq:col}
\end{equation}

\subsection{A simple example}

Here is a colour reduction algorithm $A\colon 4 \cto 3$ that reduces the number of colours from $4$ to~$3$:
\[
    A(x,1,z) = 1, \quad
    A(x,2,z) = 2, \quad
    A(x,3,z) = 3, \quad
    A(x,4,z) = \min(\{1,2,3\} \setminus \{x,z\}).
\]
For example, if we have a properly $4$-coloured path with the input colours
\[
    (1,2,1,4,3,4,3),
\]
and we apply function $A$ in all local neighbourhoods, we get a properly $3$-coloured path
\[
    (1,2,1,2,3,1,3).
\]
At the endpoints we can apply the standard trick: \emph{pretend} that the nodes near the endpoints have a neighbour of some different colour.

\subsection{Our contribution}

We present a simple approach that enables us to do iterative colour reduction e.g.\ as follows:
\[
    10^{100} \cto 12 \cto 4 \cto 3.
\]
That is, in only $3$ steps, we can reduce the number of colours from astronomical numbers to only $3$. Naturally getting below $3$ colours is not possible with any local rule; $2$-colouring paths is an inherently global problem.

We are not aware of any prior algorithm that is equally fast and that does not need to assume e.g.\ a global orientation. As we will see in the next section, our algorithm is also much simpler than prior algorithms. Our algorithm is also completely self-contained, so one can safely skip the next section.

\section{Prior work}

A typical theoretical presentation of this topic gives a number of algorithms, all of which have a complexity of $O(\log^* k)$ or $\frac{1}{2}\log^*(k) + O(1)$ iterations for a colour reduction from $k$ to~$3$. However, to highlight the differences between the algorithms, we will use here concrete numbers; our goal is to reduce the number of colours from at least $M = 2^{128}$ to~$3$ (note that $M$ is chosen so that we can use e.g.\ IPv6 addresses as colours). As we will see, the seemingly innocent $O(1)$ part actually dominates here.

Most of the prior algorithms proceed in two steps. First, we develop a colour reduction algorithm \emph{assuming} that there is a well-defined global orientation (each node has at most one predecessor and at most one successor). Put otherwise, we develop an algorithm that satisfies \eqref{eq:col} but not necessarily \eqref{eq:sym}. Then, using such an \emph{asymmetric algorithm} as a black box, we design a \emph{symmetric algorithm} that also satisfies \eqref{eq:sym}.

\subsection{Asymmetric algorithms}

The classical example of an asymmetric algorithm is the algorithm by Cole and Vishkin from 1986 \cite{cole86deterministic}. In its modern form, it is a colour reduction algorithm $2^k \cto 2k$. Iterating the rule, we get
\[
    2^{128} \cto 256 \cto 16 \cto 8 \cto 6,
\]
at which point the algorithm is stuck. At 6 colours we can switch to a naive algorithm that eliminates one colour per round, and overall we obtain
\[
    2^{128} \cto 256 \cto 16 \cto 8 \cto 6 \cto 5 \cto 4 \cto 3.
\]
In $7$ steps we reduce the number of colours from $M$ to $3$. This is where a typical introductory lecture on the topic stops.

However, we can do better. A particularly elegant approach is to resort to the algorithm by Naor and Stockmeyer \cite{naor95what}, which gives a colour reduction of $\binom{2k}{k} \cto 2k$. This rule has a fixed point at $4$ colours, after which we can do one round of the naive algorithm. Overall, we get
\[
    10^{50000} \cto 184756 \cto 20 \cto 6 \cto 4 \cto 3.
\]
Only $5$ steps from way beyond $M$ to $3$ colours.

There is also a simple trick that is applicable to both Cole--Vishkin and Naor--Stockmeyer algorithms. These algorithms only use the colour of the predecessor, and ignore the colour of the successor, that is,
\[
    A(x,y,z) = A(x',y,z) \text{ for all } x,x'.
\]
We can simulate two steps of any such algorithm in one iteration; for Cole--Vishkin we obtain
\[
    2^{128} \cto 16 \cto 6 \cto 5 \cto 4 \cto 3
\]
and for Naor--Stockmeyer we obtain
\begin{equation}
    10^{50000} \cto 20 \cto 4 \cto 3. \label{eq:ns}
\end{equation}
We are down to only $3$ iterations for colour reduction from beyond $M$ to $3$.

Here $3$ rounds is optimal for these values of the parameters. Also, the last step $4 \cto 3$ is optimal; a simple computer search shows that there is no $5 \cto 3$ algorithm \cite{rybicki11msc,rybicki15exact}. However, the step $20 \cto 4$ leaves some room for improvement; there is a $24 \cto 4$ algorithm \cite{rybicki11msc}.

\subsection{From asymmetric to symmetric algorithms}

Equipped with an efficient asymmetric algorithm, for example \eqref{eq:ns}, we can then apply it in undirected paths in a somewhat ad-hoc fashion. Here is a concrete example:
\begin{enumerate}
    \item Use the local minima and local maxima of the colours to split the path in fragments that consist of strictly increasing colours.
    \item Orient each fragment in the direction of increasing colours.
    \item Apply an efficient asymmetric algorithm to each such fragment to iteratively find a $3$-colouring. Now everything except the local minima and local maxima are properly coloured.
    \item Label the local minima with colour $4$ and local maxima with colour $5$, to obtain a proper $5$-colouring.
    \item Run two rounds of the naive colour reduction algorithm to get back to $3$ colours.
\end{enumerate}
Some care is needed to make sure we do not lose too many rounds in each step; for example, we want to do useful colour reduction already in the first round, in addition to just identifying the local minima and maxima. With a little bit of thought, combined with \eqref{eq:ns}, we obtain something along the lines of
\[
    10^{50000} \cto 20+2 \cto 4+2 \cto 3+2 \cto 4 \cto 3,
\]
where ``$+2$'' refers to the local minima and local maxima that will use two additional colours. This is $5$ rounds for $M$ to $3$. Our new algorithm will only need $3$ rounds, and it is much more streamlined.

\subsection{Symmetric algorithms by design}

There are very few efficient algorithms that are symmetric by design; most available algorithms are derived from asymmetric algorithms by following a scheme similar to the one sketched above.

However, there are some colour reduction algorithms that are designed for general undirected graphs, and we can naturally apply them also in the case of graphs of maximum degree $2$. The classical example is Linial's \cite{linial92locality,kuhn06complexity} algorithm. The algorithm is based on so-called \emph{cover-free set families} \cite{Erdos1985}, and we obtain slightly different algorithms based on the specific choice of the set family. Unfortunately, all of the constructions give at best an exponential colour reduction $k \cto O(\log k)$ per round; for example, one of the constructions used by Linial gives $k \cto 5 \lceil 4 \log k \rceil$.

Our new algorithm is much faster; it provides a doubly-exponential colour reduction $k \cto O(\log \log k)$ per round, which is asymptotically optimal for both symmetric and asymmetric algorithms \cite{linial92locality}.

\section{Our algorithm}

Fix a target number of colours $c$. We will consider
\begin{itemize}
    \item \emph{subsets} $X \subseteq [c]$,
    \item \emph{families} $\cX$ that consist of such subsets, and
    \item \emph{collections} $\ccA$ that consist of such families.
\end{itemize}
It turns out that certain collections can be directly interpreted as colour reduction algorithms; we will call such collections \emph{colourful} collections.

Section~\ref{ssec:coll} gives the definition of a colourful collection. Section~\ref{ssec:coll-to-alg} shows that any colourful collection $\ccA$ gives a colour reduction algorithm $k \cto c$ for $k = |\ccA|$. Section~\ref{ssec:alg-to-coll} shows that the converse is also true: any colour reduction algorithm $k \cto c$ gives a colourful collection $\ccA$ of size $k$. Finally, Section~\ref{ssec:cons} shows how to construct large colourful collections.

\subsection{Colourful collections}\label{ssec:coll}

We say that a collection $\ccA$ is colourful if the following holds:
\begin{enumerate}[label=(P\arabic*)]
    \item \label{p1} For each family $\cX \in \ccA$, for all subsets $X, Y \in \cX$, we have $X \cap Y \ne \emptyset$.
    \item \label{p2} For all families $\cX \ne \cY$, there are $X \in \cX$ and $Y \in \cY$ such that $X \cap Y = \emptyset$.
\end{enumerate}

\paragraph{A simple example.}

Here is an example of a colourful collection, for $c = 3$ and $|\ccA| = 4$:
\[
    \ccA = \Bigl\{
    \bigl\{\{1\}\bigr\},
    \bigl\{\{2\}\bigr\},
    \bigl\{\{3\}\bigr\},
    \bigl\{\{1,2\}, \{1,3\}, \{2,3\}\bigr\}
    \Bigr\}.
\]
To keep the examples easier to read, we leave out two innermost levels of brackets and commas and write simply
\[
    \ccA = \bigl\{
    1,\ 
    2,\ 
    3,\ 
    12\ 13\ 23
    \bigr\}.
\]

\subsection{From colourful collections to algorithms}\label{ssec:coll-to-alg}

We now show how to use a colourful collection $\ccA$ to do colour reduction $k \cto c$ for $k = |\ccA|$. The algorithm proceeds as follows. Here it will be convenient to imagine that each undirected edge $\{u,v\}$ is a pair of directed edges.

\paragraph{High-level plan:}
\begin{enumerate}
    \item Label each node $u$ with a family $\cL(u) \in \ccA$.
    \item Using the labels of the nodes $t$ and $u$, label the directed edge $(t,u)$ with a subset $L(t,u) \in \cL(u)$.
    \item Using the labels of the incoming edges $(t,u)$ and $(v,u)$, label the node $u$ with a colour $\ell(u) \in L(t,u) \cap L(v,u)$.
\end{enumerate}

\paragraph{Detailed description:}
\begin{enumerate}
    \item Apply a bijection $[k] \to \ccA$ to label each node $u$ with a family $\cL(u) \in \ccA$.
    \item Consider an edge $\{t,u\}$. By \ref{p2}, there are $X \in \cL(t)$ and $Y \in \cL(u)$ with $X \cap Y = \emptyset$. Assign the labels $L(t,u) = Y$ and $L(u,t) = X$.
    \item Consider a node $u$ with the incoming edges $(t,u)$ and $(v,u)$. Note that $L(t,u), L(v,u) \in \cL(v)$. By \ref{p1}, there is a $c \in L(t,u) \cap L(v,u)$. Set $\ell(u) = c$.
\end{enumerate}

\paragraph{Correctness.}

We already argued above that the algorithm is well-defined, assuming that $\ccA$ is colourful. Let us now show that it indeed produces a proper colouring with $c$ colours. If not, there would be an edge $\{u,v\}$ with $\ell(u) = \ell(v)$. But by construction, we have $\ell(u) \in L(v,u)$, $\ell(v) \in L(u,v)$, and $L(u,v) \cap L(v,u) = \emptyset$, a contradiction.

\subsection{From algorithms to colourful collections}\label{ssec:alg-to-coll}

We will now show that colourful collections are not only sufficient in the design of one-round colour reduction algorithms, but they are also necessary: given any colour reduction algorithm $A\colon k \cto c$, we can construct a colourful collection $\ccA$ of size $k$. We proceed as follows:
\begin{enumerate}
\item For any two colours $x,y \in [k]$, $x \ne y$, define the subset $F_{x,y} = \{ A(x,y,z) : z \ne y\}$.
\item For any colour $y \in [k]$, define the family $\cF_{y} = \{ F_{x,y} : x \ne y\}$.
\item Define the collection $\ccA = \{ \cF_{y} : y \in [k]\}$. 
\end{enumerate}

\paragraph{Correctness.}
First, consider a family $\cF_y$ for some $y$. Let $X,Y \in \cF_y$. Then there are $x \ne y \ne z$ such that $X = F_{x,y}$ and $Y = F_{z,y}$. By definition, we have $A(x,y,z) \in F_{x,y}$ and $A(z,y,x) \in F_{z,y}$, and by \eqref{eq:sym} also $A(x,y,z) = A(z,y,x)$. Hence there is a common element, $A(x,y,z)$, in $X \cap Y$, and \ref{p1} holds.

Second, let $x \ne y$, and consider the families $\cF_x$ and $\cF_y$. We have $F_{y,x} \in \cF_x$ and $F_{x,y} \in \cF_y$. If there was a common element $a \in F_{x,y} \cap F_{y,x}$, there would also exist some $s$ and $t$ with $A(x,y,s) = A(y,x,t) = A(t,x,y)$, which contradicts \eqref{eq:col}; if we had a path with the colours $(t,x,y,s)$ in this order, algorithm $A$ would fail to colour the middle nodes properly. Therefore we have $X \in \cF_x$ and $Y \in \cF_y$ with $X \cap Y = \emptyset$. This shows that \ref{p2} holds.

Together with \ref{p1}, the above argument also shows that $\cF_x \ne \cF_y$ for $x \ne y$, and hence the size of the collection is indeed $|\ccA| = k$.

\subsection{Explicit constructions of colourful collections}\label{ssec:cons}

We have already seen in Section~\ref{ssec:coll} an example of a colourful collection for $c = 3$ and $|\ccA| = 4$. Now we present the general scheme, for an even $c$. We will use the case of $c = 4$ as a running example. To simplify the notation, let $s = \binom{c}{c/2}$.
\begin{enumerate}
\item Take all subsets of size $c/2$ (there are $s$ such subsets):
\[
    \bigl\{ 12,\ 13,\ 14,\ 23,\ 24,\ 34 \bigr\}.
\]
\item Split these in $s/2$ pairs of a subset and its complement:
\[
   \bigl\{ (12, 34),\ (13, 24),\ (14, 23) \bigr\}.
\]
\item Collection $\ccA_0$ contains all $2^{s/2}$ families that we can form by picking one half of each pair:
\[
\begin{split}
\ccA_0 = \bigl\{\ 
& 12\ 13\ 14,\\
& 12\ 13\ 23,\\
& 12\ 24\ 14,\\
& 12\ 24\ 23,\\
& 34\ 13\ 14,\\
& 34\ 13\ 23,\\
& 34\ 24\ 14,\\
& 34\ 24\ 23
\ \bigr\}.
\end{split}
\]
\item Collection $A_1$ is formed from $A_0$ by augmenting each family with all subsets of size $c-1$:
\[
\begin{split}
\ccA_1 = \bigl\{\ 
& 12\ 13\ 14\ 123\ 124\ 134\ 234,\\
& 12\ 13\ 23\ 123\ 124\ 134\ 234,\\
& 12\ 24\ 14\ 123\ 124\ 134\ 234,\\
& 12\ 24\ 23\ 123\ 124\ 134\ 234,\\
& 34\ 13\ 14\ 123\ 124\ 134\ 234,\\
& 34\ 13\ 23\ 123\ 124\ 134\ 234,\\
& 34\ 24\ 14\ 123\ 124\ 134\ 234,\\
& 34\ 24\ 23\ 123\ 124\ 134\ 234
\ \bigr\}.
\end{split}
\]
\item Then form a collection $\ccA_2$ that contains all $c$ families that contain just one subset of size~$1$:
\[
\begin{split}
\ccA_2 = \bigl\{\ 
& 1,\\
& 2,\\
& 3,\\
& 4
\ \bigr\}.
\end{split}
\]
\item Finally, set $\ccA = \ccA_1 \cup \ccA_2$.
\end{enumerate}
Note that in the running example, we have $c = 4$ and $|\ccA| = 12$, and hence this is a colour reduction algorithm $12 \cto 4$, assuming that the collection is indeed colourful.

\paragraph{Correctness.}

Collection $\ccA_0$ satisfies both \ref{p1} and \ref{p2}: Consider some $\cX, \cY \in \ccA$, and let $X \in \cX$, and let $\bar X = [c] \setminus X$ be the complement of $X$. Now $\bar X$ is the only set of size $h$ with $X \cap \bar X = \emptyset$, and we have $\bar X \notin \cX$, which satisfies \ref{p1}, and $\bar X \in \cY$, which satisfies \ref{p2}.

The augmented collection $\ccA_1$ still satisfies both \ref{p1} and \ref{p2}: we only added subsets that intersect with everything, satisfying \ref{p1}, and for the purposes of \ref{p2} we can ignore the extra subsets.

The trivial collection $\ccA_2$ clearly also satisfies both \ref{p1} and \ref{p2}.

Finally, we need to argue that $\ccA$ also satisfies both \ref{p1} and \ref{p2}. Property \ref{p1} follows directly from the fact that $\ccA_1$ and $\ccA_2$ satisfy it. What remains to be checked is that \ref{p2} holds even if we pick $\cX \in \ccA_1$ and $\cY \in \ccA_2$ (or vice versa). But this is trivial, as $\cY$ contains just one subset $Y \in \cY$ of one element, and its complement is contained in every $\cX \in \ccA_1$.

\paragraph{Analysis.}

By construction, for every even $c$ we have a colour reduction scheme
\[
    2^{s/2} + c \cto c,
\]
where $s = \binom{c}{c/2}$. As a concrete example, we obtain the following three-round colour reduction scheme:
\[
    2^{462} + 12 \cto 12 \cto 4 \cto 3.
\]
Asymptotically, the scheme is doubly-exponential; we have $k \cto O(\log \log k)$.

\paragraph{Remarks.}

If we are doing colour reduction $k \cto c$, we can assign the input colours $1,2,\dotsc,c$ to the singleton families in $\ccA_2$. This way the colour reduction algorithm has an additional nice property: the nodes that already have a colour at most $c$ do not change their colours. This is a natural property, but prior efficient algorithms (e.g.\ Naor--Stockmeyer and Cole--Vishkin) do not guarantee this.

\bibliography{col-red}

\begin{thebibliography}{10}

\bibitem{barenboim13distributed}
Leonid Barenboim and Michael Elkin.
\newblock {\em Distributed Graph Coloring: Fundamentals and Recent
  Developments}.
\newblock Synthesis Lectures on Distributed Computing Theory. Morgan {\&}
  Claypool Publishers, 2013.

\bibitem{cole86deterministic}
Richard Cole and Uzi Vishkin.
\newblock Deterministic coin tossing with applications to optimal parallel list
  ranking.
\newblock {\em Information and Control}, 70(1):32--53, 1986.

\bibitem{Erdos1985}
P.~Erd{\H{o}}s, P.~Frankl, and Z.~F{\"u}redi.
\newblock Families of finite sets in which no set is covered by the union of
  {$r$} others.
\newblock {\em Israel Journal of Mathematics}, 51(1):79--89, Dec 1985.

\bibitem{kuhn06complexity}
Fabian Kuhn and Roger Wattenhofer.
\newblock On the complexity of distributed graph coloring.
\newblock In {\em Proc.\ PODC 2006}, pages 7--15, 2006.

\bibitem{linial92locality}
Nathan Linial.
\newblock Locality in distributed graph algorithms.
\newblock {\em SIAM Journal on Computing}, 21(1):193--201, 1992.

\bibitem{naor95what}
Moni Naor and Larry Stockmeyer.
\newblock What can be computed locally?
\newblock {\em SIAM Journal on Computing}, 24(6):1259--1277, 1995.

\bibitem{peleg00distributed}
David Peleg.
\newblock {\em Distributed Computing: A Locality-sensitive Approach}.
\newblock Society for Industrial and Applied Mathematics, Philadelphia, PA,
  USA, 2000.

\bibitem{rybicki11msc}
Joel Rybicki.
\newblock Exact bounds for distributed graph colouring.
\newblock Master's thesis, Department of Computer Science, University of
  Helsinki, May 2011.

\bibitem{rybicki15exact}
Joel Rybicki and Jukka Suomela.
\newblock Exact bounds for distributed graph colouring.
\newblock In {\em Proc.\ SIROCCO 2015}, pages 46--60, 2015.

\bibitem{suomela12course}
Jukka Suomela.
\newblock Distributed algorithms, 2016.
\newblock Online textbook.

\bibitem{wattenhofer13lecture}
Roger Wattenhofer.
\newblock Lecture notes on principles of distributed computing, 2017.
\newblock Online lecture notes.

\end{thebibliography}

\end{document}